\documentclass[12pt]{article}
\usepackage{graphicx,verbatim,array,multicol,amsfonts,amsmath,subfigure}
\usepackage{fullpage}
\usepackage{chicago}
\usepackage{color}
\usepackage{epsfig,latexsym,amssymb}
\usepackage{lscape}
\usepackage{bm}

\def\EndProof{{\begin{flushright}\vspace{-2mm}$\Box$\end{flushright}}}

\newtheorem{theo}{Theorem}[section]

\begin{document}

\title{On sequential Monte Carlo, partial rejection control and approximate Bayesian computation}
\author{
    G. W.~Peters\footnote{School of Mathematics and Statistics, University of New South Wales, Sydney, Australia} \footnote{CSIRO  Sydney, Locked Bag 17, North Ryde, NSW, 1670, Australia}
    \and  Y. Fan$^*$
    \and S. A.~Sisson$^*$\footnote{Communicating author: Scott.Sisson@unsw.edu.au}
}
\maketitle

\begin{abstract}
We present a sequential Monte Carlo sampler variant of the partial
rejection control algorithm, 
and show that this variant can be
considered 
as a sequential Monte Carlo
sampler 
with a modified mutation kernel.
We prove that the new sampler
can reduce 
the variance of the incremental importance weights when compared with standard sequential Monte Carlo samplers. 
We provide 
a study of theoretical properties of the new algorithm, and make connections with some existing algorithms. Finally, the sampler is 
adapted for application under the challenging ``likelihood free," approximate Bayesian computation modelling framework,
where we demonstrate superior performance over existing likelihood-free samplers.

\vspace{0.5cm} \noindent \textbf{Keywords:} Approximate Bayesian
computation; Bayesian computation; 
Likelihood-free inference; Sequential Monte Carlo samplers;
Partial rejection control.

\end{abstract}

\section{Introduction}
\label{sec:intro}

Sequential Monte Carlo (SMC)  methods have  emerged out of
the fields of engineering, probability and statistics in recent years. 
Variants of the methods sometimes appear under the names of particle filtering or interacting 
particle systems (e.g. \shortciteNP{Arulampalam};
\shortciteNP{Andrieu}; \shortciteNP{delmoral04};
\shortciteNP{doucet+dg01}), and their theoretical properties have been extensively studied  (\shortciteNP{Crisan}; \shortciteNP{delmoral04}; \shortciteNP{Kunsch}).

The standard SMC algorithm involves finding a numerical solution to a set of
filtering recursions, such as filtering problems arising from non-linear / non-Guassian state
space models.  Under this framework, the SMC algorithm samples from a
(often naturally occurring) sequence of distributions $\pi_{t}$, indexed by $t=1,\ldots,T$.  
Each distribution is defined on the support $E^{t}=E\times E\times ...\times E$.  
\shortciteN{delmoral+dj06} (see also \shortciteNP{Peters05})
generalize the SMC algorithm to the case where the distributions $\pi_t$ are all defined on the 
same support $E$. This generalization, termed the SMC {\it sampler},  adapts the SMC algorithm to
the more popular setting in which the state space $E$ remains static.

In short, the SMC sampler generates weighted samples (termed {\it particles}) from a sequence of distributions $\pi_t$,
for $t=1,\ldots, T$, where $\pi_T$ may be of particular interest.  
We refer to $\pi_T$ as the target distribution. Procedurally,  particles obtained from an arbitrary initial distribution $\pi_1$, with a set of corresponding initial weights,  are sequentially propagated through each distribution  $\pi_t$ in the sequence via three processes, involving mutation (or move), correction (or importance weighting) and selection (or resampling).
The final weighted particles at distribution $\pi_T$ are considered weighted samples from the target distribution $\pi$. The mechanism is similar to sequential importance sampling 
(resampling)
 (\shortciteNP{liu01}; \shortciteNP{doucet+dg01}), with one of the crucial differences being the framework under which the particles are allowed to move, resulting in differences in the calculation of the weights of the particles.
 
One of the major difficulties with SMC-type algorithms is particle depletion, in which the
weights of the majority of the particles gradually decrease to zero, while a few particle weights dominate the population. 
This severely increases the variability of  Monte Carlo estimates of expectations under $\pi$.
In this article, we develop an algorithm which incorporates the partial rejection control (PRC) strategy of \shortciteN{liu01} into the SMC sampler framework. A particular motivation for this stems from the recent developments in ``likelihood-free'' (or approximate Bayesian) computation \shortcite{beaumont+zb02,marjoram+mpt03,sisson+ft07},  where an extremely high proportion of mutated particles are expected to have very 
small, or exactly zero,  posterior weights.

In this article, we develop the SMC samplers PRC algorithm, in which the partial rejection
control mechanism is built directly into the mutation kernel of the SMC sampler. In this manner, a  particle mutation may be rejected if the resulting importance weight is below a certain threshold. 
We begin with a brief introduction to the standard sequential Monte Carlo sampler in Section \ref{sec:SMC}, and then present the SMC sampler PRC algorithm. We also
discuss implementational issues arising from the inclusion of the PRC stage, including estimation for the resultant kernel normalizing constant.
Section \ref{sec:analysis} provides some theoretical results that justify the addition of PRC in terms of improvements in the variance of the incremental importance weights. 
We also discuss a central limit theorem and derive
a recursive expression for the asymptotic variance of our algorithm. 
In addition, we make a novel connection between the
SMC sampler PRC algorithm and the AliveSMC algorithm from the 
rare-event literature developed in \citeN{LeGland2}. In Section \ref{sec:ABC}
we adapt the SMC sampler PRC algorithm for application in the likelihood-free
modelling framework, and demonstrate the computational gains achieved over existing likelihood-free algorithms via a simulated example. Finally, we present a stochastic claims reserving analysis using the developed methods
in Section \ref{sec:examples}, and  conclude with a discussion.

\section{Sequential Monte Carlo and partial rejection}
\label{sec:SMC}

\subsection{Sequential Monte Carlo sampler}

\shortciteN{delmoral+dj06} introduced a modification of the sequential Monte Carlo algorithm,
termed the sequential Monte Carlo {\it sampler}. Consider a sequence of distributions
$\pi_t(x),  t=1,\ldots, T$, with $x \in E$, where the final distribution $\pi_T$ is the distribution of interest. By introducing
a sequence of
backward kernels $L_k$,  a new distribution 
$\widetilde{ \pi }_{t} (x_1, \ldots, x_t ) =\pi _{t} ( x_{t})
\prod\limits_{k=1}^{t-1}L_{k}\left( x_{k+1},x_{k}\right)$ may be defined for the {\it path} of a particle
$(x_1, \ldots, x_t) \in E^t$ through the sequence $\pi_1,\ldots,\pi_t$. 
The only restriction on the backward kernels is that the correct marginal distributions 
$\int \widetilde{\pi} _{t}( x_1, \ldots, x_t ) dx_1, \ldots, dx_{t-1} = \pi_t(x_t)$ 
are available.

Within this framework,
one may then work with the sequence of distributions, $\widetilde{ \pi }_t$,
under the standard SMC algorithm.  
In summary, the SMC sampler algorithm involves three stages: {\it mutation}, whereby the particles
are moved from $x_{t-1}$ to $x_t$ via a mutation kernel $M_t(x_{t-1},x_t)$ as described below (\shortciteNP{Johansen09}; \shortciteNP{delmoral+dj06});
{\it correction},  where the particles are reweighted with respect to $\pi_t$ via the incremental importance weight (\ref{eqn:weight}); and {\it selection}, where according to some measure of particle diversity, commonly the effective sample size (ESS, 
\shortciteNP{Kunsch}; \shortciteNP{Kitigawa};
\shortciteNP{doucet+dg01}; \shortciteNP{liu+c98}), the weighted particles may be resampled in order to reduce the variability of the importance weights. 

In more detail, suppose that at
time $t-1$, the distribution $\widetilde{ \pi }_{t-1}$ can be
approximated empirically by $\widetilde{ \pi }_{t-1}^{N}$
using $N$ weighted particles. These particles are first propagated to the next distribution
$\widetilde{ \pi }_{t}$ using a mutation kernel $M_t(x_{t-1},x_t)$, and then assigned new
weights $W_t = W_{t-1}w_{t}\left( x_1, \ldots x_t\right)$, where
$W_{t-1}$ is the weight of a particle at time $t-1$ and $w_{t}$ is the incremental weight
given by
\begin{equation}\label{eqn:weight}
w_{t}\left( x_1,\ldots, x_t\right) =\frac{\widetilde{\pi }_{t}\left(
x_1,\ldots, x_t\right) }{\widetilde{\pi }_{t-1}\left( x_1, \ldots, x_{t-1}\right)
M_{t}\left( x_{t-1},x_{t}\right) }=\frac{\pi _{t}\left(
x_{t}\right) L_{t-1}\left( x_{t},x_{t-1}\right) }{\pi _{t-1}\left(
x_{t-1}\right) M_{t}\left( x_{t-1},x_{t}\right) }.
\end{equation}
The resulting particles are now weighted samples from   $\widetilde{ \pi }_{t}$.
Consequently from (\ref{eqn:weight}), under the SMC sampler framework, one may work directly with the marginal distributions $\pi_t(x_t)$ such that $w_t(x_1,\ldots,x_t)=w_t(x_{t-1},x_t)$. While the choice of the backward kernels $L_{t-1}$ is essentially arbitrary, 
their specification can strongly affect the
performance of the algorithm. See \shortciteN{delmoral+dj06} for detailed discussion.

\subsection{Incorporating partial rejection control}
\label{sec:SMCPRC}

It is well known that the performance of SMC methods are strongly
dependent on the mutation kernel \shortcite{cornebise+mo08}. If $M_t$ is poorly chosen, 
such that it does not place particles in regions of the
support of $\pi _{t}$ with high density, then many importance
sampling weights will be close to zero. This leads to
sample degeneracy, as a few well located particles with large weights dominate the particle
population, resulting in large variance for estimates made using these samples. 

\citeN{liu01} (see also
\shortciteNP{liu}) 
introduced the partial rejection control strategy to overcome
particle degeneracy in a sequential importance sampling setting. Under this mechanism, when
the weight of a particle at distribution $\pi_t$ falls below a finite threshold, $c_t\geq 0$, the particle
is probabilistically discarded. It is replaced with a particle drawn from the previous distribution $\pi_{t-1}$ which is then mutated to $\pi_t$. This new particle's weight is then compared to the threshold, with this process repeating until a particle is accepted.
This approach is  termed {\it partial} rejection, as the replacement particle is drawn from $\pi_{t-1}$, not $\pi_1$ \shortcite{liu}.

Under the SMC sampler framework 
we modify this approach and incorporate the partial rejection mechanism directly within the mutation kernel. Hence at time $t-1$, the particle $x_{t-1}$ is moved via the mutation kernel $M_t(x_{t-1}, x_t)$ and weighted according to (\ref{eqn:weight}). This particle is accepted with probability 
$p$, determined by the particle's weight and the weight threshold $c_t$. If rejected,
a new particle is obtained via the mutation kernel $M_t$, until a particle is accepted. 

For the sequence of distributions $\pi_t$, $t=1,\ldots,T$, the mutation and backward kernels $M_t$ and $L_{t-1}$, a sequence of weight thresholds $c_t$, and  PRC normalizing constants $r(c_t, x_{t-1})$ (defined below), the SMC sampler PRC algorithm is given by:

\noindent{\bf SMC sampler PRC algorithm}
\vskip 0.3cm
\hrule width 430pt depth 1pt
\vskip 0.4cm
\begin{description}
\item[Initialization:] 
Set $t=1$. \\For $i=1,\ldots, N$, sample $x^{(i)}_1\sim \pi_1(x)$,  and set
weights $W_1(x^{(i)}_1) = \frac{1}{N}$.

\item[Resample:] Normalize the weights $\sum_i W_t(x_t^{(i)})=1$. If $[\sum_i W_t(x_t^{(i)})^2]^{-1}<H$ 
resample $N$ particles with respect to $W_t(x_t^{(i)})$ and set $W_t(x_t^{(i)})=\frac{1}{N}, i=1,\ldots,N$. 

\item[Mutation and correction:] Set $t=t+1$ and $ i=1$:\\
\begin{tabular}{ll}
(a)& Sample  $x_t^{(i)}\sim M_t(x_{t-1}^{(i)},x_t)$ and set weight for 
$x_t^{(i)}$ to  \\
       &  \quad\quad $W_t(x_t^{(i)})=W_{t-1}(x^{(i)}_{t-1})\frac{\pi_{t}(x_t^{(i)}) L_{t-1}( x_t^{(i)},x_{t-1}^{(i)})}{\pi_{t-1}(x_{t-1}^{(i)}) M_t(x_{t-1}^{(i)},x_t^{(i)})}$.\\
      (b)& With probability $1-p^{(i)}=1-\min\{ 1,W_t(x_t^{(i)})/c_t\}$, reject $x_t^{(i)}$ and go to (a). \\
      (c)&  Otherwise, accept $x_t^{(i)}$ and set  $W_t(x_t^{(i)}) =  W_t(x_t^{(i)})r(c_t,x_{t-1}^{(i)})/p^{(i)}$.\\
      (d)&       Increment $i=i+1$. If $i\leq N$, go to (a). \\   
       (e)&       If $t< T$ go to Resample.\\
\end{tabular}
\end{description}
\vskip 0.0cm
\hrule width 430pt depth 1pt
\vskip 0.5cm

The above algorithm without the mutation and correction steps (b) and (c) is equivalent to the standard SMC sampler algorithm \shortcite{delmoral+dj06}.
In the resample stage, the degeneracy of the particle approximation is quantified through the usual estimate of the effective sample size, $1\leq[\sum_i W_t(x_t^{(i)})^2]^{-1}\leq N$ \cite{liu+c98}. We discuss the choice of the thresholds, $c_t$, in later Sections. 

The addition of a rejection step at each time $t$ effectively modifies
the mutation kernel $M_t$. We denote the resulting kernel by $M_t^{\ast },$ where
\begin{equation}
\label{eqn:emstar}
	M_{t}^{\ast }\left( x_{t-1},x_{t}\right) =r(c_{t}, x_{t-1})^{-1} \min \left[ \left\{ 1, W_{t-1}(x_{t-1})\frac{w_t\left(
	x_{t-1},x_{t}\right) }{c_{t}}\right\} M_{t}\left(
	x_{t-1},x_{t}\right)\right].
\end{equation}
The quantity $r(c_{t}, x_{t-1})$ denotes the normalizing constant for particle $x_{t-1}$, given by
\begin{equation}\label{Rejection Probability}
r(c_{t}, x_{t-1})=\int \min \left\{ 1, W_{t-1}(x_{t-1})\frac{w_t\left(
x_{t-1},x_{t}\right) }{c_{t}}\right\} M_{t}\left(
x_{t-1},x_{t}\right) dx_{t}.
\end{equation}
Note that $0<r(c_t,x_{t-1})\leq 1$ if  (w.l.o.g.) the mutation kernel $M_t$ is normalized, so that $\int M_{t}(x_{t-1},x_t)dx_t=1$, and if  the PRC threshold $0\leq c_t<\infty$ is finite.
Thus the SMC sampler PRC algorithm can be considered as an
SMC sampler algorithm with the mutation kernel $M_{t}^{\ast
}\left( x_{t-1},x_{t}\right)$, and the correction weight
\begin{equation}\label{eqn:weightPRC}
W_{t}(x_{t})
=
W_{t-1}(x_{t-1})\frac{\pi _{t}\left( x_{t}\right)
L_{t-1}\left( x_{t},x_{t-1}\right) }{ \pi _{t-1}\left( x_{t-1}\right)
M_{t}^{\ast }\left( x_{t-1},x_{t}\right) }.
\end{equation}

\subsection{Estimation of the normalizing constant}
\label{NoNorm}

As the normalizing constant $r(c_t, x_{t-1})$ in the weight calculation (\ref{eqn:weightPRC}) in general depends on $x_{t-1}$, it must be evaluated. Where no analytic 
solution can be found, approximating (\ref{Rejection Probability}) may be
achieved 
by, for example, quadrature
methods if the sample space $E$ is relatively low dimensional or
Monte Carlo methods if $E$ is high dimensional. For example, for $j=1,\ldots,m$ independent samples
$x_t^{*(j)}$ sampled from  $M_{t}\left(x_{t-1},x_{t}\right)$
$$
\hat{r}(c_{t}, x_{t-1}) \approx \frac{1}{m}\sum_{j=1}^m \min \left\{ 1, W_{t-1}(x_{t-1})\frac{w\left(
x_{t-1},x^{*(j)}_{t}\right) }{c_{t}}\right\}.
$$
An alternative, computationally more efficient approach is to select kernels $M_t$ and $L_{t-1}$ such
that  $r(c_t,x_{t-1})=r(c_t)$ will be constant for all particles $x_{t-1}$. 
In this case, the value of $r(c_t)$ may be absorbed into the proportionality constant of the weights, and safely ignored.
Equation (\ref{Rejection Probability})  suggests that this can be achieved if  $M_t(x_{t-1},x_t)$, $W_{t-1}(x_{t-1})$ and $w(x_{t-1},x_t)$ are independent of $
x_{t-1}$.

Specifying mutation kernels $M_t$ such that $M_t(x_{t-1},x_t)=M_t(x_t)$ amounts
to choosing a {\it global} kernel
which is the same for all particles $x_{t-1}$. This is common in practice (e.g. \citeNP{west93}).
The particle dependent weight $W_{t-1}(x_{t-1})$ can be set to $1/N$ for all particles following a resampling (or preselection) step. Finally, consider for a moment the backward kernel of the form
\begin{equation}
\label{eqn:L3}
    L_{t-1}^{opt}(x_t,x_{t-1})=\frac{\pi_{t-1}(x_{t-1})M_t(x_{t-1},x_t)}
        {\int\pi_{t-1}(x_{t-1})M_t(x_{t-1},x_t)dx_{t-1}}.
\end{equation}
This backward kernel is an approximation of the optimal backward kernel, in the sense of
 the choice of $L_{t-1}$ that 
minimizes the variance of the importance sampling weights \shortcite{delmoral+dj06}.
Under the backward kernel (\ref{eqn:L3}), the incremental weight can be approximated by
\begin{eqnarray}
\label{eqn:L3noConst}
    w_t(x_{t-1},x_t)
    & = &
    \pi_t(x_t)/\int \pi_{t-1}(x_{t-1})M_t(x_{t-1},x_t)dx_{t-1}\nonumber\\
    & \approx &
    \pi_t(x_t)/\sum_{i=1}^N W_{t-1}(x^{(i)}_{t-1})M_t(x^{(i)}_{t-1},x_t) 
    \nonumber.
\end{eqnarray}
Under a global mutation kernel $M_t(x_t)$,  and following a resampling step, 
then the incremental weight under
this backward kernel reduces to
$w_t(x_{t-1},x_t)= \pi_t(x_t)/M_t(x_t) $, which is independent of $x_{t-1}$. Thus, the weight
calculation in 
(\ref{eqn:weightPRC}) becomes
\begin{eqnarray*}
W_t(x_t) &\propto& \pi_t(x_t)/\left[\min \left\{1, \frac{w(x_{t-1},x_t)}{Nc_t}\right\}M_t(x_t)\right]\\
& = &\left\{\begin{array}{ll}\pi_t(x_t)/M_t(x_t) & \mbox{ if } \min \left\{ 1, \frac{w\left(
x_{t-1},x_{t}\right) }{Nc_{t}}\right\}=1\\Nc_t& \mbox{ otherwise}.\end{array}\right.
\end{eqnarray*}
Note that under this setting, the SMC sampler PRC  algorithm can be considered as a sequence of importance sampling strategies with partial rejection control.

\section{SMC Sampler PRC algorithm analysis}
\label{sec:analysis}

In this section we study theoretical properties of the SMC sampler PRC algorithm.
We firstly bound the variance of the importance weights, 
and then present a central limit theorem for the
sampler
with a recursive expression for the asymptotic variance.
Finally, via a connection
with an existing algorithm, we establish a condition for which the number of rejection steps under the PRC mechanism is almost surely finite.

\subsection{Variance of the incremental weights}
\label{sec:AV}

We begin this section by establishing a bound on the variance of the importance weights of the SMC sampler PRC algorithm.

\begin{theo}
\label{theorem1} 
Let $W_{t}\left( x_{t}\right)$
denote the importance sampling weight at time $t$ from a standard SMC sampler with mutation kernel $M_t$,
and 
let $W_{t}^{\ast }\left( x_{t}\right)$ denote the equivalent 
weight following a partial rejection control step under the SMC sampler PRC algorithm, with resulting mutation kernel $M_t^{\ast}$.
Then
$$\mathbb{V}ar_{M_{t}^{\ast }}\left[
W_{t}^{\ast }\left( x_{t}\right) \right] \leq
\mathbb{V}ar_{M_{t}}\left[ W_{t}\left( x_{t}\right) \right].$$
\end{theo}
\noindent{\bf Proof:} See Appendix A.1.

Hence, applying  partial rejection control within the SMC sampler framework will not
worsen, and may improve the variance of the importance weights, by reducing the $\chi^2$-distance between the sampling and target distributions at each stage, $t$. 

%


In the case where $\min \left\{
1,\frac{W_t\left(x_{t}\right) }{c_{t}}\right\} =1$ for all $x_t$, which is
achieved when $c_{t} \leq \inf_{x_{t}}\left\{ W_t\left(
x_{t}\right)\right\}$, then from (\ref{Rejection Probability}) we have $r(c_t,x_{t-1})=1$ for all $x_{t-1}$.
From (\ref{eqn:emstar}), this
results in $M_t^{\ast}(x_{t-1},x_t)=M_t(x_{t-1},x_t)$ and hence
$\mathbb{V}ar_{M_{t}^{\ast
}}\left[ W_{t}^{\ast }\left(x_{t}\right) \right] =\mathbb{V}%
ar_{M_{t}}\left[ W_{t}\left(x_{t}\right) \right]$. That
is, the SMC sampler PRC algorithm
reduces to the standard SMC sampler when $c_{t} \leq \inf_{x_{t}}\left\{ W_t\left(
x_{t}\right)\right\}$, and in this case, the variance of the importance weights is maximised. 
When $W_t(x_t)\in[0,\infty)$ this is realized for $c_t=0$ where we define $0/0:=1$.

\subsection{A central limit theorem}

Central Limit Theorems (CLTs) for 
SMC and particle filtering algorithms have been derived in various
literatures \shortcite{delmoral04,Kunsch,chopin2,delmoral+dj06,johansen2008nap}.
They are based on the observation
that an SMC algorithm introduces local errors (fluctuations)
as a result of the approximations introduced by
sampling numerically from the transitions. Hence, at each stage $t$, one can
decompose the error between the 
target distribution $\pi_t$ and the 
$N$-particle 
approximation $\pi_t^N$. This turns out to be a sum of the local sampling
fluctuations at each discrete time in the past,
propagated forwards in time to 
$t$.

In the setting of the SMC sampler algorithm, the existence of a CLT is established
by
\shortciteN{delmoral04}. Explicitly,  under the assumption of multinomial
resampling at each stage of the algorithm,
and the integrability conditions given in \shortciteN{chopin2} [Theorem 1] and
\shortciteN{delmoral04} [Section 9.4, pp.300-306], then 
for 
a suitable continuous and bounded test function $\varphi \in
C_{b}\left( E\right)$ we have
\begin{equation}
\label{eqn:clt}
	N^{1/2}\left(\mathbb{E}_{\pi_t^N}(\varphi)-\mathbb{E}_{\pi_t}(\varphi)\right)
	\rightarrow
	{\mathcal N}\left(0, V_{SMC,t}(\varphi)\right)
\end{equation}
as $N\rightarrow\infty$, for each $t=1,\ldots,T$.
\shortciteN{delmoral+dj06} obtain a 
recursive expression for the asymptotic variance $V_{SMC,t}(\varphi)$
as an explicit function of the
backward kernels $L_{t-1}$ and the sequence of distributions on path space, $\widetilde{\pi}_t$.

Following \shortciteN{delmoral+dj06}, we obtain an analogous result
for the SMC sampler PRC algorithm. Under the same assumptions as the above, we have the CLT (\ref{eqn:clt}) with asymptotic variance given by
\begin{equation}\label{eqn:AVsmcprc}\begin{array}{lll}
&&V _{SMC-PRC,t}\left( \varphi \right) = \int I_{1}\frac{\widetilde{\pi }%
_{t}^{2}\left( x_{1}\right) }{\pi _{1}\left( x_{1}\right) }\left(
\int \varphi \left( x_{1}\right) \widetilde{\pi }_{1}\left(
x_{t}|x_{1}\right) dx_{t}-\mathbb{E}_{\pi _{t}}\left( \varphi
\right) \right) ^{2}dx_{1}   \\
&&+\sum\limits_{k=2}^{t-1}\int I_{k}\frac{\left( \widetilde{\pi
}_{t}\left( x_{k}\right) L_{k-1}\left( x_{k},x_{k-1}\right)
\right) ^{2}}{{\pi }_{k-1}\left(
x_{k-1}\right)M_k\left(x_{k-1},x_k\right)}\left( \int \varphi
\left( x_{t}\right) \widetilde{\pi }_{t}\left( x_{t}|x_{k}\right) dx_{t}-%
\mathbb{E}_{\pi _{t}}\left( \varphi \right) \right) ^{2}dx_{k-1} dx_k   \\
&&+\int I_{t}\frac{\left( \pi _{t}\left( x_{t}\right)
L_{t-1}\left( x_{t},x_{t-1}\right) \right) ^{2}}{\pi _{t-1}\left(
x_{t-1}\right) M_t(x_t,x_{t-1})}\left( \varphi \left( x_{t}\right)
-\mathbb{E}_{\pi _{t}}\left( \varphi \right) \right)
^{2}dx_{t-1}dx_t  
\end{array}
\end{equation}
with $I_{k} = \left[ r(c_{k}, x_{k-1})^{-1}\min \left\{ 1
,\frac{1}{N}\frac{\pi _{k}\left( x_{k}\right) L_{k-1}\left(
x_{k},x_{k-1}\right) }{\pi _{k-1}\left( x_{k-1}\right)
M_k\left(x_{k-1},x_k\right)c_{k}}\right\} \right] ^{-1}$ and
$I_{1}=1. $ 

The contribution of the PRC stage to the asymptotic variance is encapsulated in the $I_k$ terms.
Under the standard SMC sampler algorithm we have $c_k\leq\inf_{x_t}\left\{W_t(x_t)\right\}$ 
so that $I_k=1$ for all $k=1,\ldots,t$. In this setting, (\ref{eqn:AVsmcprc}) reduces to the asymptotic variance expression obtained by \shortciteN{delmoral+dj06}.

\subsection{Connections to an existing SMC algorithm}
\label{sec:connect}

In rare event applications there is a high probability of generating particles with exactly zero weights. The AliveSMC algorithm \shortcite{LeGland2,LeGland} was developed to ensure that a particle population of 
a desired size
persists at each iteration of a standard SMC algorithm
(see \shortciteNP{DelMoral,johansen+dd06} for related methods).
In this setting, 
the number of particles at each time $t$ is considered as a random variable $N_{c_t}$. That is, $N_{c_t}$ is the number of particles required to generate exactly $N$ non-zero weighted particles.
In this Section we
reinterpret the SMC sampler PRC algorithm in terms of the AliveSMC algorithm. As a consequence, in Section \ref{sec:finite} we are able to establish  a condition under which the PRC resampling stage will require a finite number of rejection attempts.

In
\shortciteN{LeGland}, at iteration $t$, a fitness function
is applied to select particles
satisfying a desired criteria. Those particles not satisfying the criteria receive a zero weight. In an SMC sampler PRC
setting, the fitness function can be interpreted as
selecting those particles
with a weight that is immediately accepted under the PRC acceptance probability.
As such, 
we may rewrite the SMC sampler PRC algorithm with a modified mutation and correction step:

\newpage
\noindent{\bf Reinterpreted SMC sampler PRC/AliveSMC algorithm}
\vskip 0.3cm
\hrule width 430pt depth 1pt
\vskip 0.4cm
\begin{description}

\item[Mutation and correction:] Set $t=t+1$ and $j=1$. 
For $ i=1,\ldots, N_{c_t}$:\\
\begin{tabular}{ll}
(a)& Sample  $x_t^{(i)}\sim M_t(x_{t-1}^{(j)},x_t)$ and calculate  
       $W=W_{t-1}(x_{t-1}^{(j)})\frac{\pi_{t}(x_t^{(i)}) L_{t-1}( x_t^{(i)},x_{t-1}^{(j)})}{\pi_{t-1}(x_{t-1}^{(j)}) M_t(x_{t-1}^{(j)},x_t^{(i)})}$.\\
 (b) &    Set weight for $x_{t}^{( i)}$\ as \\
  & $W_{t}(x_t^{(i)}) \propto \left\{
\begin{array}{lll} & Wr(c_t,x_{t-1}^{(j)})/p^{(i)} & \mbox{with probability } p^{(i)}=\min\{1, W/c_t\}\\
& 0 & \mbox{otherwise.}
\end{array}
\right. $ \\
(c) & If $W_t(x_t^{(i)})\neq 0$, increment $j=j+1$.\\
\vspace{-5mm}
\end{tabular}
If $t<T$ go to Resample.
\end{description}
\vskip 0.0cm
\hrule width 430pt depth 1pt
\vskip 0.5cm




Note that $j=1,\ldots,N$ indexes the particles $x_{t-1}^{(j)}$ at time $t-1$  such that particle mutations from $x_{t-1}^{(j)}$ generate a non-zero weight exactly once.
%
Also, under the fitness function, particle $x_t^{(i)}$
has a probability $1-p^{(i)}=1-\min\{ 1, W/c_t\}$
of being exactly zero.
Given that it is possible to express the SMC sampler PRC algorithm within the AliveSMC framework,
we may
adapt the results of \shortciteN{LeGland2} and \shortciteN{LeGland}, to obtain
a condition under which the SMC sampler PRC algorithm is guaranteed
to require a finite number of attempts, $N_{c_{t}}<\infty$, to obtain exactly $
N$ non-zero weighted particles.

\subsection{Analysis of the number of rejection attempts}
\label{sec:finite}

Following \shortciteN{LeGland2} and \shortciteN{LeGland}, we define the random
variable $N_{c_{t}}$ as 
\[
	N_{c_{t}} \triangleq \inf \left\{ N^*\geqslant 1:\sum\limits_{i=1}^{N^*}
	W_{t}^{\left( i\right) }(x_t)\geqslant N\sup_{x_t\in E}
	W_{t}(x_t) \right\} 
\]
\shortciteN{LeGland}
proved for the AliveSMC algorithm that the random number of
particles $N_{c_{t}}$ is almost surely finite with
$N_{c_{t}}\geqslant N$,
under the condition that $\left\langle \pi_{t-1}M_{t},W_{t}\right \rangle  = 
\frac{\int W_t(x_t) \int \pi_{t-1}(x_{t-1})M_t(x_{t-1}, x_t) dx_{t-1} dx_t}{\int \pi_{t-1}(x_{t-1})M_t(x_{t-1}, x_t)   dx_{t-1}}
>0$.
A sufficient condition for this to
hold is $\mathbb{E}_{M_{t}}\left[ W_{t}\left( x_{t}\right)
\mid x_{t-1}=x\right] >0, \label{Condition1}$ for all $x \in E$. Thus
for the SMC sampler PRC algorithm, we have the following theorem:

\begin{theo}
\label{theorem3} Under the SMC sampler PRC algorithm, the number
of rejection attempts at each stage of the algorithm, 
$N_{c_{t}}\geqslant N$,
is almost surely finite
if $c_t<\infty$.
\end{theo}
\noindent{\bf Proof:} See Appendix A.2.

\noindent \textbf{Corollary 5.1} \textit{The following convergence in
probability holds, with a rate of $1/\sqrt{N}$:
\begin{equation*}
\frac{N_{c_t}}{N}\rightarrow \frac{\sup_{x_t\in E}W_t(x_t)
}{\left\langle \pi_{t-1}M_{t},W_{t}\right\rangle }
<\infty .
\end{equation*}
}See \shortciteN{LeGland} for further details.
Hence, the SMC sampler PRC algorithm possesses an almost surely finite number of rejection attempts if the PRC threshold $c_t$ is finite, with the above rate of convergence.

\section{Approximate Bayesian computation}
\label{sec:ABC}
With the aim of posterior simulation from $\pi( x| {\cal D})\propto\pi({\cal D}|x)\pi(x)$ for parameters $x$ and observed data ${\cal D}$,  ``likelihood-free,"
approximate Bayesian computation (ABC) 
methods
 are often utilised 
 when the likelihood function, $\pi({\cal D}|x)$, is computationally intractable or when its evaluation is computationally prohibitive.
ABC methods can be based on rejection sampling
\shortcite{tavare+bgd97,beaumont+zb02}, 
Markov chain Monte Carlo 
\shortcite{marjoram+mpt03,bortot+cs07,ratmann+ahwr09}
and SMC-type samplers \shortcite{sisson+ft07,toni+wsis08,beaumont+cmr08,delmoral+dj08}. 
While currently among the most efficient ABC methods, the underlying practical issue with SMC-type algorithms is in avoiding sample degeneracy through extreme numbers of particles with low or exactly zero weights.
In this Section, we will demonstrate that the SMC sampler PRC algorithm applied within the ABC framework can achieve significant performance gains and greater modelling flexibility over existing SMC-type ABC methods.

The underlying approach of ABC methods is to augment the (intractable) posterior to $\pi(x,{\cal D}'|{\cal D})\propto\pi({\cal D}|{\cal D}',x)\pi({\cal D}'|x)\pi(x)$, where the auxiliary parameter is an artificial data set distributed according to the model ${\cal D}'\sim\pi(\cdot|x)$. An approximation of the target posterior $\pi(x|{\cal D})$ is then given by
\begin{equation}
\label{eqn:abcapproxn}
	\pi_{ABC}(x|{\cal D}) 
	\propto 
	\int \pi({\cal D}|{\cal D}', x)\pi({\cal D}'|x)\pi(x) d{\cal D}' .
\end{equation}
The weighting distribution $\pi({\cal D}|{\cal D}',x)$ takes high density in regions where the datasets ${\mathcal D}$ and ${\cal D}'$ are similar, and low density otherwise. Comparison of the datasets is usually achieved through low-dimensional summary statistics $T(\cdot)$, so that, for example
\begin{equation}\label{eqn:indicator}
	\pi({\cal D}|{\cal D}',x)\propto\left\{\begin{array}{cl}1&\mbox{if }\rho(T({\cal D}),T({\cal D}'))\leq\epsilon\\0&\mbox{else},\end{array}\right.
\end{equation}
for some small tolerance value $\epsilon>0$ and distance measure $\rho$.
If $T(\cdot)$ are sufficient statistics, and $\epsilon\rightarrow 0$ so that $\pi({\cal D}|{\cal D}',x)$ reduces to a point mass at $T({\cal D})=T({\cal D}')$ then $\pi_{ABC}(x|{\cal D})=\pi(x|{\cal D})$ is recovered exactly, otherwise the ABC approximation to $\pi(x|{\cal D})$ is of the form (\ref{eqn:abcapproxn}), with greater accuracy for smaller $\epsilon$. The computational overhead of all ABC samplers increases as $\epsilon$ decreases, producing a trade off between computation and accuracy.
ABC methods either sample from the joint density $\pi(x,{\cal D}'|{\cal D})$ by arranging to cancel out the intractable likelihood in a weight or acceptance probability, or sample from $\pi_{ABC}(x|{\cal D})$ directly via Monte Carlo integration
\begin{equation}
\label{eqn:abc-monte-carlo}
	\pi_{ABC}(x|{\cal D}) \approx
	\frac{\pi(x)}{S}\sum_{s=1}^S \pi({\cal D}|{\cal D}'_{s},x),
\end{equation}
where ${\cal D}'_1,\ldots,{\cal D}'_S\sim\pi({\cal D}'|x)$ are draws from the likelihood given $x$.  Almost all current ABC methods have the weighting density 
(\ref{eqn:indicator}) written directly into the algorithm.

We apply the SMC sampler PRC algorithm in the ABC framework as follows:
The target $\pi_t(x_t)=\pi_{ABC,t}(x_t|{\cal D})$ is given by (\ref{eqn:abcapproxn}), with the weighting function $\pi_t({\cal D}|{\cal D}',x)$ parameterized by a different scaling parameter $\epsilon_t$ for each $t$, where $\infty=\epsilon_1\geq\ldots\geq\epsilon_T$, produces increasing accuracy at each step, $t$.  The $\epsilon_t$ sequence and its length, $T$, may be determined  {\it a priori} or dynamically. Evaluation of $\pi_t(x_t)$ is defined by (\ref{eqn:abc-monte-carlo}) through $S$ Monte Carlo draws. 
Given the high computational overheads of ABC methods, we 
avoid evaluating the PRC normalizing constant  (as demonstrated in Section \ref{NoNorm}), through a global mutation kernel $M_t(x_t)$, the backward kernel $L^{opt}_{t-1}$ (c.f. \ref{eqn:L3}) and enforced resampling.

\noindent{\bf SMC sampler PRC-ABC algorithm}
\vskip 0.3cm
\hrule width 430pt depth 1pt
\vskip 0.4cm
\begin{description}
\item[Initialization:] 
Set $t=1$. \\For $i=1,\ldots, N$, sample $x^{(i)}_1 \sim \mu(x)$,  and set
weights 
$W_t(x^{(i)}_1) = \pi_{ABC,1}(x_1^{(i)}|{\cal D})/\mu(x^{(i)}_1)$.

\item[Resample:] Resample $N$ particles with respect to $W_t(x_t^{(i)})$ and set $W_t(x_t^{(i)})=\frac{1}{N},\\ i=1,\ldots,N$.

\item[Mutation and correction:] Set $t=t+1$ and $ i=1$:\\
\begin{tabular}{ll}
(a)& Sample  $x_t^{(i)}\sim M_t(x_t)$ and set weight for 
$x_t^{(i)}$ to  \\
&  \quad\quad $W_t(x_t^{(i)})=\pi_{ABC,t}(x_t^{(i)} | {\cal D}) /M_t(x_t^{(i)})$.\\
(b)& With probability $1-p^{(i)}=1-\min\{ 1,W_t(x_t^{(i)})/c_t\}$, reject $x_t^{(i)}$ and go to (a). \\
(c)& Otherwise, accept  $x_t^{(i)}$ and set  $W_t(x_t^{(i)}) =  W_t(x_t^{(i)})/p^{(i)}$.\\
(d)&    Increment $i=i+1$.   If $i\leq N$, go to (a).\\
(e)&  If  $t<T$ then go to Resample.
\end{tabular}
\end{description}
\vskip 0.0cm
\hrule width 430pt depth 1pt
\vskip 0.5cm

The density $\mu(x)$ is an initial sampling distribution, from which direct sampling is available. 
As with the tolerance $\epsilon_t$, the PRC thresholds $c_t$ may also be determined dynamically (see below for an illustration). 
Note that as the resampled particles in the Resample step play no subsequent  part in the sampler, in practice this step can be omitted. The path of each particle $(x_1^{(i)},\ldots,x_T^{(i)})\in E^T$ can be reconstructed post-simulation, if required, by resampling the recorded marginal populations $(W_t(x_t^{(i)}), x_t^{(i)})$.

The above algorithm has a number of benefits over existing SMC-type ABC samplers \shortcite{sisson+ft07,toni+wsis08,beaumont+cmr08,delmoral+dj08}. Firstly, the weighting density $\pi({\cal D}|{\cal D}',x)$ can take any form -- we suggest any smoothing kernel, following \citeN{blum09}. Existing samplers in the literature are restricted to the uniform function (\ref{eqn:indicator}). Secondly, there is complete control over the PRC threshold, $c_t$, unlike \shortciteN{sisson+ft07} who impose a specific value. Thirdly, in estimating $\pi_{ABC}(x_t|{\cal D})$, as long as the $L_{t-1}^{opt}$ backward kernel is used, any number $S\geq 1$ of Monte Carlo draws can be used in (\ref{eqn:abc-monte-carlo}). Existing samplers only use $S=1$, and so there is less control over the variability of the weights. Finally, providing that the computation required to estimate the PRC normalizing constants, $r(c_t,x_{t-1})$, is acceptable, a form of the SMC sampler PRC-ABC sampler may be constructed which uses arbitrary mutation and backward kernels, allowing the user to select the most appropriate tools for a given problem.

\subsection{Simulation study}
\label{section:simstudyabc}

We now demonstrate the superior performance of the SMC sampler PRC-ABC algorithm through a controlled study.
Specifically, we specify the true posterior  $\pi(x|{\cal D})$ as $N(0,1)$ by defining the likelihood and prior as ${\cal D}\sim N(x,1)$ and $\pi(x)\propto 1$, with a single observed datum, ${\cal D}=0$.
For this model, a sufficient statistic is $T({\cal D})={\cal D}$. 
From (\ref{eqn:abcapproxn}), for the uniform weighting density (\ref{eqn:indicator}) with $\rho(a,b)=|a-b|$ and $\epsilon=\epsilon_u$, or for  $\pi({\cal D'}\mid {\cal D}, x) = N({\cal D}, \epsilon_g^2)$, then $\pi_{ABC}(x|{\cal D})$ may be obtained in closed form as
\[ 
	\pi_{ABC}(x|{\cal D})\propto\frac{\Phi(\epsilon_u-x)-\Phi(-\epsilon_u-x)}{2\epsilon_u}
	\qquad\mbox{or}\qquad
	\pi_{ABC}(x|{\cal D})=N(0,1+\epsilon_g^2)
\]
respectively, where $\Phi(\cdot)$ denotes the standard Gaussian CDF. 
In both cases $\pi_{ABC}(x|{\cal D})\rightarrow N(0,1)$ as $\epsilon\rightarrow 0$. 
In order to directly compare the two approximate posteriors we impose equal variances on the two weighting functions, so that $\epsilon_g=\sqrt{3}\epsilon_u$

We adopt the following sampler specifications:
A particle population of size $N=1000$ was drawn from the initial sampling distribution $\mu(x)\sim U(-5,5)$, and the sequence of distributions, $\pi_1,\ldots,\pi_{10}$, is defined by
$
	\{\epsilon_t\}=\{\infty,10,5,2,1,0.5,0.2,0.1,0.05,0.05\},
$
on the $\epsilon_g$ scale.
The mutation kernel $M_t(x_t)=\sum_{j=1}^NW_{t-1}^{(j)}\psi(x_t|x_{t-1}^{(j)},\tau^2)$ is taken as a Normal kernel density estimate of $\pi_{t-1}(x_{t-1})$, with $\tau^2=1$ and where $\psi(x|\mu,\sigma^2)$ denotes the PDF of a $N(\mu,\sigma^2)$ distribution evaluated at $x$.
We initially use $S=1$  Monte Carlo
draws to approximate $\pi_{ABC}(x|{\cal D})$  (c.f. \ref{eqn:abc-monte-carlo}).

\begin{figure}[htb]
\begin{center}
\includegraphics[width=10cm,angle=-90]{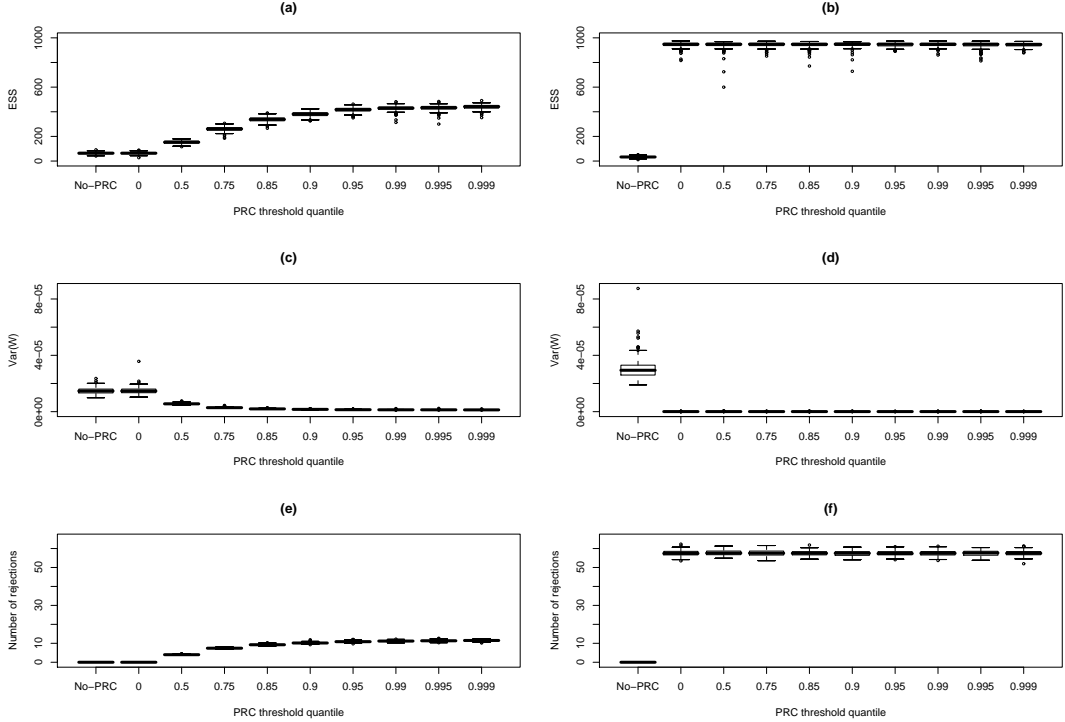}\\
\caption{\small\label{fig:sim-abc} Effective sample size (a,b), variance of normalized importance weights (c,d) and mean number of rejections per particle (e,f) as functions of PRC threshold $c_t$. PRC threshold is defined dynamically as a quantile of the non-zero weights at time $t$ (x-axis). Left plots (a,c,e) and right plots (b,d,f) are obtained under the Gaussian and uniform weighting densities $\pi({\cal D}|{\cal D}',x)$ respectively. Boxplots are based on 250 sampler replications.
}
\end{center}
\end{figure}

Figure \ref{fig:sim-abc} examines the effect of PRC on the effective sample size (ESS), the variance of the importance weights and the mean number of rejections per particle. The PRC threshold was determined dynamically at each iteration as $c_t=Q(W_t^+(x_t),q)$, the $q$-th quantile of the {\it non-zero} weights at time $t$ (obtained by mutating all $x_{t-1}$ particles under $M_t$ before implementing the PRC stage), for $q=0,0.5, 0.75, 0.85, 0.9, 0.95, 0.99, 0.995, 0.999$. 
Results are shown using the Gaussian (left plots) and uniform (right plots) weighting density, based on 250 sampler replications.
Note that the PRC threshold with $q=0$ approximately corresponds to a standard SMC sampler (``No-PRC") only for the Gaussian weighting function, as the uniform weighting function permits exactly zero importance weights.
Setting $q=0$ for the uniform weighting density corresponds to existing SMC-type ABC samplers \shortcite{sisson+ft07,toni+wsis08,beaumont+cmr08}.

For both weighting densities, the effective sample size increases as $c_t$ increases, and the variance of the importance weights decreases.
Naturally, the higher the PRC threshold, the more rejections occur, quantifying the extra computation required for the gains in sampler performance.
However, there is a notable difference in the transition from poor (no PRC) to improved (under PRC) performance between the two different weighting densities. This occurs as the uniform weighting density only permits $0/1$ weights, compared to the smoother scale under the Gaussian. As a result, the uniform weighting density (which is the only choice under existing ABC samplers) has a fixed, albeit strong, performance gain over not implementing PRC, but at a very high computational cost (Figure \ref{fig:sim-abc},f). Comparison with panels (a,c,e) suggests that considerable computational gains can be achieved with alternative weighting functions, without sacrificing sampler performance. This is easily permitted under the SMC sampler PRC framework.

\begin{figure}[bth]
\begin{center}
\includegraphics[width=10cm,angle=-90]{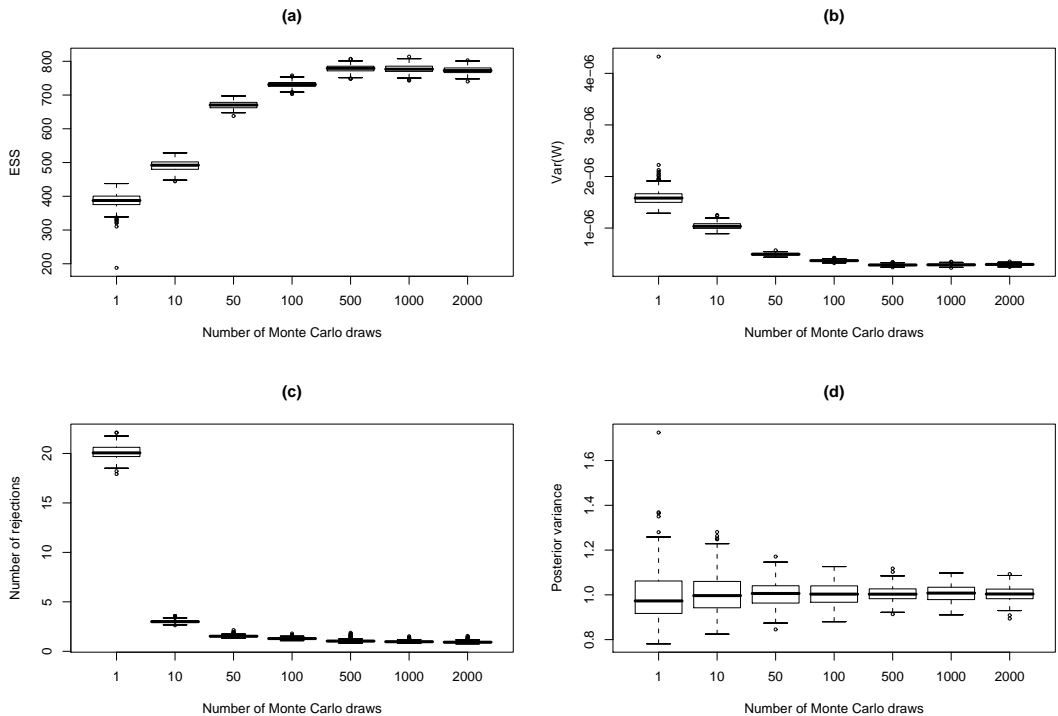}\\
\caption{\small\label{fig:abc-s} The effect of the number of Monte Carlo draws, $S$, on sampler performance. Panels show (a) effective sample size (ESS), (b) variance of the importance weights, (c) the mean number of rejection attempts, and (d) estimates of the posterior variance (true value $\approx 1$), as a function of the number of Monte Carlo draws in the estimation of $\pi_{ABC}(x|{\cal D})$.
}
\end{center}
\end{figure}

When using the $L_{t-1}^{opt}$ backward kernel (\ref{eqn:L3}), any number $S\geq 1$ of Monte Carlo draws may be used to approximate $\pi_{ABC}(x|{\cal D})$ via (\ref{eqn:abc-monte-carlo}). While $S=1$ is near universal under existing ABC algorithms, one would expect to realize less variable importance weights for $S>1$.
Figure (\ref{fig:abc-s}) illustrates the effect of increasing $S$, under the Gaussian density function, based on the PRC threshold $c_t=Q(W_t^+(x_t),0.95)$.
An increase in the effective sample size (panel a) is reflected by the reduction in the variability of the importance weights (panel b), as is the variability in the estimates of the posterior variance (panel d).  This in turn results in lower numbers of rejections at the PRC stage (panel c). Of course, these performance gains are again balanced by the strong increases in computation required for $S>1$. It would appear that unless the data-generation procedure ${\cal D}_s'\sim \pi({\cal D}|x)$ is computationally inexpensive, $1\leq S\leq 10$ would seem to be the most useful choice in practice. Regardless, the greatest gains in sampler performance under the SMC sampler PRC algorithm are achieved for $S=1$.

\section{A stochastic claims reserving analysis}
\label{sec:examples} 

We present an analysis of an important and popular
class of statistical models in actuarial science using
stochastic claims reserving. 
We consider a time series formulation of the distribution-free chain
ladder model (\citeNP{mack93}; \citeNP{gisler+w08}; \shortciteNP{peters+ws08}).
For a claim
on an insurance company for an accident in year $i$, $C_{i,j}$
denotes the cumulative claim in subsequent years $j\geq i$.
Cumulative claims can refer to payments, claims incurred and other expenses. At
time $I$, we have observations $\mathcal{D}_{I}=\left\{
C_{i,j};~i+j\leq I\right\}$, and for reserving against future
claims we wish to predict $\mathcal{D}_{I}^{c}=\left\{
C_{i,j};~i+j>I,~i\leq I\right\}$. 
One such dataset is illustrated in Table \ref{table:triangle}.

Under a time series formulation, cumulative claims $C_{i,j}$ in
different accident years $i$ are independent and satisfy, for $j=0,\ldots,I-1$,
\begin{equation}
\label{eqn:dfcl}
    C_{i,j+1}=f_{j}C_{i,j}+\sigma _{j}\sqrt{C_{i,j}}\varepsilon_{i,j+1},
\end{equation}
where $\mathbf{f}=\left(f_{0},\ldots, f_{I-1}\right)$ and
$\bm{\sigma} =\left( \sigma_{0},\ldots,\sigma_{I-1}\right)$ are
respectively the chain ladder factors and standard deviations, and
the residuals $\varepsilon _{i,j}$ are i.i.d. with mean 0 and
variance 1. The model  is constrained such that
$P(C_{i,j}>0|\{C_{k,0}\}_{k=1}^j,\mathbf{f},\bm{\sigma})=1$ for
all $i,j$ (see \shortciteNP{peters+ws08}).
If distributional assumptions are made on the residuals
$\varepsilon_{i,j}$
(e.g. \citeNP{yao08}), the
posterior distribution can be made computationally tractable.
However, a primary intention of this model is to work with distribution-free assumptions on the residuals, and therefore on the cumulative claims.
Within this distribution-free context one wishes to quantify
popular 
risk metrics such as value-at-risk and expected-shortfall
to be calculated for the predicted claims distribution, both
of which are highly relevant to regulatory reporting. 
Alternative approaches,
based on credibility results,
can relax such distributional
assumptions, but can only provide statements on posterior first and second
moments in limiting cases \cite{gisler+w08}.

Previously, actuaries have proceeded by predicting claims via a
deterministic model known as the classical chain ladder algorithm.
This approach predicts unobserved future cumulative claims by the
recursion
 $\widehat{C}_{i,I-i}={C}_{i,I-i}$, and for $j>I-i$
\begin{equation}
\label{eqn:classical-cl}
    \widehat{C}_{i,j}=\widehat{C}_{i,j-1}\widehat{f}^{(CL)}_{j-1}
    \qquad\mbox{where}\qquad
    \widehat{f}^{(CL)}_{j-1}=\frac{\sum\nolimits_{i=0}^{I-j}C_{i,j}}{%
    \sum\nolimits_{i=0}^{I-j}C_{i,j-1}},
\end{equation}
and where, in the time series formulation, the variances are
estimated by
\[
    \widehat{\sigma }_{j}^{2(CL) } =
    \frac{1}{I-j-1}\sum\nolimits_{i=0}^{I-j-1}C_{i,j}\left( \frac{C_{i,j+1}}{%
    C_{i,j}}-\widehat{f}_{j}^{(CL) }\right) ^{2}.
\]
See \citeN{mack93} for an estimator of
$\widehat{\sigma}_{I-1}^{2(CL)}$. As this algorithm is
deterministic there is strong interest in stochastic chain ladder
models, which naturally allow the quantification of uncertainty,
such as the mean square error of prediction. In the claims
reserving setting the most popular stochastic models are those
with estimators which recover the classical chain ladder
estimators. We consider one such Bayesian stochastic model which
has the property that as the diffusivity of the priors
$\pi(\mathbf{f},\bm{\sigma})$ tends to infinity
$\widehat{\mathbf{f}}^{(MMSE)} \rightarrow
\widehat{\mathbf{f}}^{(CL)}$ where $MMSE$ denotes the posterior
mean \cite{gisler+w08}. Hence by (\ref{eqn:classical-cl}) the
posterior mean $E[C_{i,J}|\mathcal{D}_I]=\widehat{C}_{i,J}$
recovers the classical estimators, thereby justifying the
classical model. 
We sample from the
intractable posterior
$\pi_{ABC}(\mathbf{f},\bm{\sigma}|\mathcal{D}_I)$ using the SMC sampler
PRC-ABC algorithm.

\subsection{Analysis and results}

This model is interesting as the intractability of the likelihood
directly impacts the ability to generate synthetic data sets,
$\mathcal{D}'_I$, from the model. That is, if the distributional
form  of the residuals were known, data-generation from the model would be trivial.
To retain a distribution-free setting we alternatively utilise a
conditional bootstrap approach \shortcite{peters+ws08}.
Conditional upon proposed parameters $\mathbf{f}$ and
$\bm{\sigma}$, the residuals
$\widetilde{\epsilon}_{i,j}|\mathbf{f},\bm{\sigma}$ are
iteratively obtained by inversion of (\ref{eqn:dfcl}). Then, by
independently drawing resampled residuals from the empirical
conditional residual distribution, a bootstrap sample of the
cumulative claims $\mathcal{D}'_I$ is then available through
recursion on (\ref{eqn:dfcl}).

In analyzing the real claims reserving data in Table
\ref{table:triangle} 
we
specify independent priors $f_j\sim\mbox{Gamma}(\alpha_j,\beta_j)$
with mean $\alpha_j\beta_j=\widehat{f}^{(CL)}_j$ and
$\sigma_j\sim\mbox{IGamma}(a_j,b_j)$ with mean
$b_j/(a_j-1)=\widehat{\sigma}^{(CL)}_j$ for $j=0,\ldots,I-1$, each
with large variance. For summary statistics we adopt
$T(\mathcal{D}')=(\mathcal{D}'_I,\mu'(\widetilde{\epsilon}),s'(\widetilde{\epsilon}))$
where $\mu'(\widetilde{\epsilon})$ and
$s'(\widetilde{\epsilon})$ denote the sample mean and standard
deviation of the conditionally resampled residuals
$\widetilde{\epsilon}'_{i,j}|\mathbf{f},\bm{\sigma}$. The
observed summary statistics are given by
$T(\mathcal{D})=(\mathcal{D}_I,0,1)$ following the zero mean and unit
variance assumptions on the true residuals.

We implement the SMC sampler PRC-ABC algorithm with uniform
weighting density
(\ref{eqn:indicator}) and
$\rho(T(\mathcal{D}_I),T(\mathcal{D}'_I))=[(T(\mathcal{D}_I)-T(\mathcal{D}'_I))^{\top}\Sigma^{-1}(T(\mathcal{D}_I)-T(\mathcal{D}'_I))]^{1/2}$
defined
as Mahalanobis distance,
where the covariance $\Sigma$ 
is estimated following \shortciteN{peters+ws08}.  We use $N=5000$ particles, PRC threshold
$c_t=Q(W^+_t(x_t),0)$
and a deterministic
distribution schedule $\{\epsilon_{t}\}=\{\infty,10, \ldots, 0.00001\}$ with $T=22$.
The mutation kernel
$M_t(x_t)=\sum_{i=1}^NW_{t-1}^{(i)}(x_{t-1}^{(i)})\mbox{Gamma}(a(x_{t-1}^{(i)}),b(x_{t-1}^{(i)}))$
is a mixture of gamma densities, with mean
$a(x_{t-1}^{(i)})b(x_{t-1}^{(i)})=x_{t-1}^{(i)}$
and large variance.

Table \ref{table:dfcl-predict} presents a comparison of the
parameter estimates $\widehat{\mathbf{f}}$ and
$\widehat{\bm{\sigma}}$, and predicted cumulative claims,
$\widehat{C}_{i,j}$ under classical and Bayesian models.
Given the uninformative priors, the posterior
mean estimates and resulting predicted claims agree well with
those obtained under the classical model. This provides some validation for the
deterministic classical model estimates
under the Bayesian stochastic interpretation.

\begin{figure}[htb]
\begin{center}
\includegraphics[width=10cm,angle=-90]{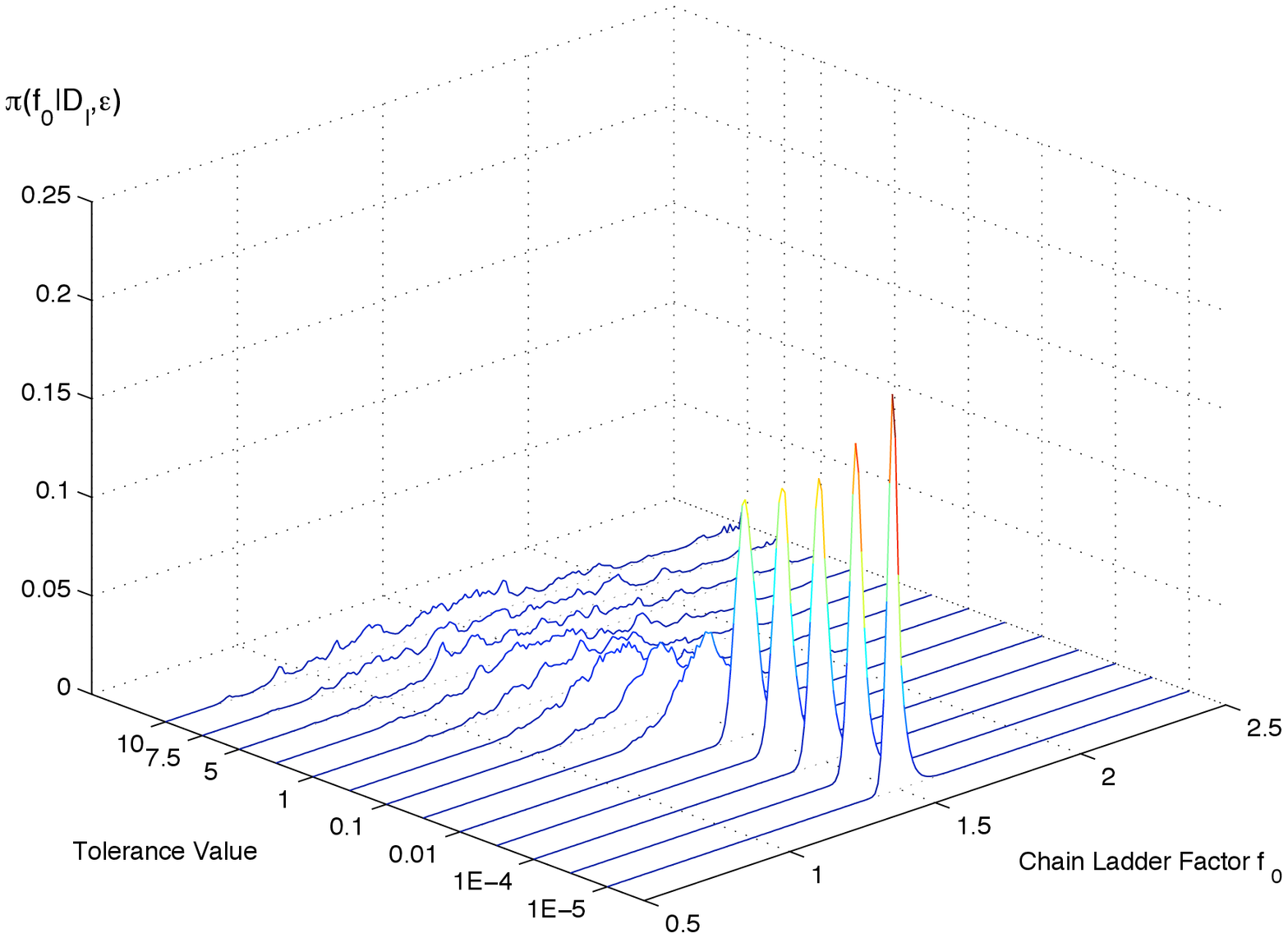}
\includegraphics[width=10cm,angle=-90]{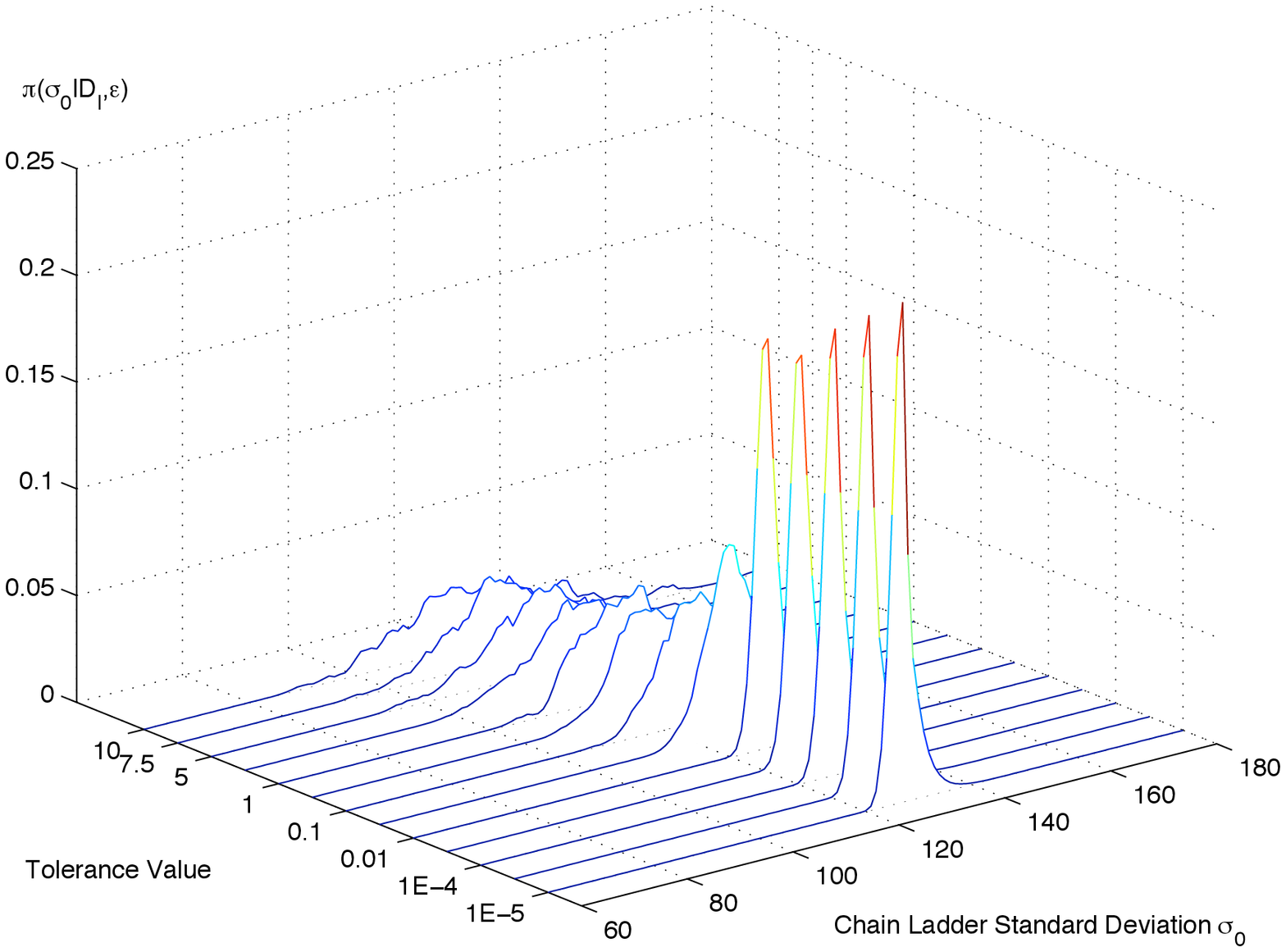}
\caption{\small\label{fig:factor-variance} Evolution of the
marginal posterior density estimates of the chain ladder factor
$\pi_t(f_0|\mathcal{D}_I)$ (left) and the associated standard deviation
$\pi_t(\sigma_0|\mathcal{D}_I)$ (right) as a function of $\epsilon_t$.}
\end{center}
\end{figure}

Perhaps more usefully for inference, the full posterior $\pi_{ABC}(\mathbf{f},\bm{\sigma}|\mathcal{D}_I)$ is 
available.
Figure \ref{fig:factor-variance}
illustrates how the estimated marginal densities of the first
chain ladder factor, $\pi_{ABC,t}(f_0|\mathcal{D}_I)$, and the
associated standard deviation, $\pi_{ABC,t}(\sigma_0|\mathcal{D}_I)$, evolve as 
$\epsilon_t$ decreases. The precision of the densities
clearly improves, as decreasing $\epsilon_t$ imposes stricter
restrictions on the permissible deviations of the ABC approximate posterior
$\pi_{ABC}(\mathbf{f},\bm{\sigma}|\mathcal{D})$ from the target posterior $\pi(\mathbf{f},\bm{\sigma}|\mathcal{D})$. A
full predictive analysis may now follow, including upper and lower
credible bounds on predicted future claims.

\section{Discussion}
\label{sec:discussion} 

When used in challenging settings, sequential Monte Carlo samplers often suffer from severe particle degeneracy.
In this article we have provided a practical approach to tackling this problem, by incorporating the partial rejection control mechanism of \shortciteN{liu01} directly into the mutation kernel of the SMC sampler. The resulting sampler will not worsen, and can improve the variance of the importance weights, sometimes substantially so.
By establishing clear relationships with existing samplers \shortcite{delmoral+dj06,LeGland2}, many theoretical properties may be extended to the SMC sampler PRC algorithm, including a central limit theorem, and a proof of an almost sure finite number of PRC rejection attempts.

There is much opportunity for the specification of the sequence of PRC thresholds to be further automated, if desired. For example, by dynamically determining $c_t>c_{t-1}$ if the effective sample size of the particle population at time $t-1$ falls too low, and conversely allowing $c_t<c_{t-1}$ if the level of PRC resampling is too high, in order to reduce computational overheads.

As the SMC sampler PRC algorithm allows practical inference in challenging situations in which particle weights are highly variable,
we anticipate that a primary application of the sampler will be within the rapidly developing ``likelihood-free'' approximate Bayesian computation framework.
The presented sampler is more flexible and efficient than existing SMC-type ABC samplers, allowing a previously unavailable 
degree of control over the 
computation utilised for a given analysis. Perhaps more importantly, the extra flexibility achieved by allowing arbitrary weighting densities (unlike existing ABC samplers) enables the analysis of improved models within the ABC framework, in line with recent non-parametric interpretations 
\cite{blum09}.

\section{Acknowledgements}

YF and SAS are supported by the Australian Research Council
through the Discovery Project scheme (DP0664970, DP0877432).
GWP is supported by an APA scholarship
and by the School of Mathematics and
Statistics, UNSW and CSIRO CMIS. The authors thank P.
Shevchenko , A. Johansen and M. Briers for thoughtful discussions.

\nocite{wuthrich+m08}
\bibliographystyle{chicago}
\bibliography{smcprcabc}

\section*{Appendix}

\subsubsection*{A.1: Proof of Theorem \ref{theorem1}}

The proof follows the arguments presented by \shortciteN{liu}.
In particular we study the SMC sampler PRC algorithm in terms of $\chi^2$ distance
between the sampling distribution and the target distribution at
stage $t$. Let
{\small
\begin{equation*}
W_{t}\left( x_{t}\right) \propto W_{t-1}(x_{t-1}) \frac{\pi
_{t}\left( x_{t}\right) L_{t-1}\left( x_{t},x_{t-1}\right) }{\pi
_{t-1}\left( x_{t-1}\right) M_{t}\left( x_{t-1},x_{t}\right) }
\quad\mbox{and}\quad 
W_{t}^{\ast }\left( x_{t}\right) \propto W_{t-1}(x_{t-1})\frac{\pi
_{t}\left( x_{t}\right) L_{t-1}\left( x_{t},x_{t-1}\right) }{\pi
_{t-1}\left(
x_{t-1}\right) M_{t}^{\ast }\left( x_{t-1},x_{t}\right) }.
\end{equation*}
}
\noindent Recall that the normalising constant for the mutation kernel $M_t^{\ast}$ at time $t$ is 
\begin{equation*}
r({c_{t}, x_{t-1}})= \int \min \left\{ 1,\frac{W_{t}(x_{t})
}{c_{t}}
\right\} M_{t}\left( x_{t-1},x_{t}\right) dx_{t} 
=\frac{1}{c_{t}}\mathbb{E}%
_{M_{t}}\left[ \min \left\{ c_{t},W_{t}(x_{t}) \right\} %
\right].
\end{equation*}
The variance of the importance weight at time $t$ from a standard SMC sampler, with respect to $%
M_{t}$, is given by
$$
\mathbb{V}ar_{M_{t}}[W_{t}(x_{t})]=\int [W_{t}(x_t)]^2M_{t}\left( x_{t-1},x_{t}\right)  dx_t - \mu^2, 
$$
and similarly,
the variance of the equivalent importance weight at time $t$ following a PRC step under the SMC sampler PRC algorithm, with respect to $%
M_{t}^{\ast }$, is given by
$$
\mathbb{V}ar_{M_{t}^{\ast }}[W^{\ast}_{t}(x_{t})]=\int [W_{t}^{\ast}(x_t)]^2M_{t}^{\ast }\left( x_{t-1},x_{t}\right)  dx_t - \mu^2,
$$
where
$$
\mu=\mathbb{E}_{M^*_t}[W^*_t(x_t) ] = \mathbb{E}_{M_t}[W_t(x_t)].
$$

\noindent We also have that
\begin{equation*}
\begin{array}{lll}
\int [W_{t}^{\ast}(x_t)]^2M_{t}^{\ast }\left( x_{t-1},x_{t}\right)  dx_t &=&
\int \left[W_{t-1}(x_{t-1})\frac{\pi(x_t)L_{t-1}(x_t, x_{t-1})}{\pi_{t-1}(x_{t-1})
M_{t}^{\ast }\left( x_{t-1},x_{t}\right)} \right]^2 M_{t}^{\ast }\left( x_{t-1},x_{t}\right)dx_t\\
&=&r(c_{t}, x_{t-1})\int\frac{W^2_{t-1}(x_{t-1})}{\min\{1 , \frac{W_t(x_t)}{c_t}\}}
 \frac{\pi^2_t(x_t)L_{t-1}^2(x_{t-1})}{\pi_{t-1}^2(x_{t-1})M^2_t( x_{t-1},x_{t})}M_t( x_{t-1},x_{t}) dx_t\\
&=&r(c_{t}, x_{t-1})\int \max \{ W^2_t(x_t),c_t W_t(x_t) \}M_t( x_{t-1},x_{t}) dx_t\\
&=&r(c_{t}, x_{t-1})\mathbb{E}_{M_t}\left[\max \{ W_t(x_t),c_t \}W_t(x_t) \right]\\
&=&\frac{1}{c_t}\mathbb{E}_{M_t}\left[ \min \left\{ c_{t},W_t(x_t) \right\}\right]\mathbb{E}_{M_t}\left[ \max \{W_t(x_t),c_t  \}W_t(x_t) \right]\\
&\leq& \frac{1}{c_t}\mathbb{E}_{M_t} \left[\min \left\{ c_{t},W_t(x_t) \right\} \max \{W_t(x_t),c_t  \}W_t(x_t) \right]\\
&=& \frac{1}{c_t}\mathbb{E}_{M_t} \left[ c_tW^2_t(x_t)\right]=\mathbb{E}_{M_t} \left[ W^2_t(x_t)\right].
\end{array}
\end{equation*}
The above inequality holds since the random variables $\min \left\{ c_{t},W_t(x_t) \right\}$ and
$\max \{W_t(x_t),c_t  \}W_t(x_t)$ are positively correlated (see \shortciteNP{liu01}),  and so
\begin{equation*}
\begin{array}{ll}
\mathbb{E}_{M_t} \left[\min \left\{ c_{t},W_t(x_t) \right\} \max \{W_t(x_t),c_t  \}W_t(x_t) \right]\\
 \quad \quad - \mathbb{E}_{M_t}\left[ \min \left\{ c_{t},W_t(x_t) \right\}\right]\mathbb{E}_{M_t}\left[ \max \{W_t(x_t),c_t  \}W_t(x_t) \right]  \geq 0.
\end{array}
\end{equation*}
Hence
$$
\mathbb{V}ar_{M_{t}^{\ast }}[W^{\ast}_{t}(x_{t})] \leq \mathbb{E}_{M_t} \left[ W^2_t(x_t) \right]-\mu^2=\mathbb{V}ar_{M_{t}}[W_{t}(x_{t})].
$$
\EndProof

\subsubsection*{A.2: Proof of Theorem \ref{theorem3}}

To satisfy the condition 
$\mathbb{E}_{M_{t}}\left[ W_{t}\left( x_{t}\right)
\mid x_{t-1}=x\right] >0, \forall x\in E,$ for the SMC sampler PRC algorithm, we require
\begin{equation}
\label{eqn:split}
	 \int_E
	 W_{t-1}(x)
	 w_t(x,x_t)
	 r(c_t,x)
	\left[
	\min \left\{ 1,
	\frac{W_{t-1}(x)
	w_t\left( x,x_{t}\right) }{c_{t}}\right\} %
	\right] ^{-1}M_{t}\left( x,x_{t}\right) dx_{t} >0 . 
\end{equation}
For a particle $x_{t-1}=x$, the proposed
state $x_t$ can take values in a support which can be
split into two regions, $A(x)$ and $A^c(x)$, such that $A(x) \cup A^c(x)=E$. 
These respectively correspond to when $\min \left\{ 1,\frac{W_{t-1}(x)w_t\left( x,x_t\right)
}{c_t}\right\} =1$ (i.e. particle acceptance probability under PRC is 1) and when
$\min
\left\{ 1,\frac{W_{t-1}(x)w_t\left( x,x_{t}\right) }{c_{t}}\right\}
=W_{t-1}(x)w_t\left( x,x_{t}\right)/c_t$ (i.e. the particle may be rejected under PRC).
Note that in the extreme cases, 
$c_t\leq W_{t-1}(x)w_t(x,x_t) \Rightarrow A^c(x)=\emptyset$ 
reduces to the SMC sampler algorithm, and $c_t>\sup_{x_t}W_{t-1}(x)w_t(x,x_t)\Rightarrow A(x)=\emptyset$.
More generally, (\ref{eqn:split}) may be expanded as
\[
	\frac{r(c_t,x)W_{t-1}(x)}{\pi _{t-1}\left( x\right)} \int_{A(x)} 
	\pi_{t}\left(x_{t}\right) L_{t-1} \left( x_{t},x\right) dx_{t} + c_tr(c_t,x)\int_{A^c(x)} M_t(x,x_t)dx_t >0
\]
which is always greater than zero for finite $c_t<\infty$ as $0<r(c_t,x)\leq 1$.
\EndProof

{\footnotesize
\begin{table}[ptbh]
\begin{center}
{\footnotesize {\scriptsize {\
\begin{tabular}{c|cccccccccc}
Accident & \multicolumn{10}{c}{Development Year, $j$}\\
{Year, $i$} & {$0$} & {$1$} & {$2$} & {$3$} & {$4$} & {$5$} & {$6$} & {$7$} & {$8$%
} & {$9$} \\ \hline
{$0$} & {$594.6975$} & {$372.1236$} & {$89.5717$} & {$20.7760$} & {$20.6704 $%
} & {$6.2124$} & {$6.5813$} & {$1.4850$} & {$1.1130$} & \multicolumn{1}{c|}{$%
1.5813$} \\ \cline{11-11} {$1$} & {$634.6756$} & {$324.6406$} &
{$72.3222$} & {$15.1797$} & {$6.7824$} & {$3.6603$} & {$5.2752$} &
{$1.1186$} & {$1.1646$} & \multicolumn{1}{|c|}{}
\\ \cline{10-10}
{$2$} & {$626.9090$} & {$297.6223$} & {$84.7053$} & {$26.2768$} & {$15.2703 $%
} & {$6.5444$} & {$5.3545$} & {$0.8924$} & \multicolumn{1}{|c}{} &
\multicolumn{1}{c|}{} \\ \cline{9-9}
{$3$} & {$586.3015$} & {$268.3224$} & {$72.2532$} & {$19.0653$} & {$13.2976 $%
} & {$8.8340$} & {$4.3329$} & \multicolumn{1}{|c}{} &  &
\multicolumn{1}{c|}{ } \\ \cline{8-8}
{$4$} & {$577.8885$} & {$274.5229$} & {$65.3894$} & {$27.3395$} & {$23.0288 $%
} & {$10.5224$} & \multicolumn{1}{|c}{} &  &  & \multicolumn{1}{c|}{} \\
\cline{7-7}
{$5$} & {$618.4793$} & {$282.8338$} & {$57.2765$} & {\ $24.4899$} & {$%
10.4957 $} & \multicolumn{1}{|c}{} &  &  &  & \multicolumn{1}{c|}{} \\
\cline{6-6} {$6$} & {$560.0184$} & {$289.3207$} & {$56.3114$} & {\
$22.5517$} & \multicolumn{1}{|c}{} &  &  &  &  &
\multicolumn{1}{c|}{} \\ \cline{5-5} {$7$} & {$528.8066$} &
{$244.0103$} & {$52.8043$} & \multicolumn{1}{|c}{} & &
\multicolumn{3}{c}{to be predicted ${Y}_{i,j}$}  &  &
\multicolumn{1}{c|}{} \\ \cline{4-4} {$8$} & {$529.0793$} &
{$235.7936$} & \multicolumn{1}{|c}{} &  &  &  &  &  & &
\multicolumn{1}{c|}{} \\ \cline{3-3} {$9$} & {$567.5568$} &
\multicolumn{1}{|c}{} &  &  &  &  &  &  &  & \multicolumn{1}{c|}{}
\\ \cline{2-11}
\end{tabular}
} }  }
\end{center}
\caption{\small A claims development triangle. Upper triangle
denotes observed annual claims $Y_{i,j}$ from which
$C_{i,j}=\sum_{k=0}^jY_{i,k}\in\mathcal{D} _{I}$ may be obtained;
lower triangle denotes annual $Y_{i,j}$ and cumulative claims
${C}_{i,j}\in\mathcal{D}_{I}^{c}$ to be predicted. Data are real
insurance figures in units of \$10,000 (c.f. W\"uthrich and Merz,
2008). The triangle assumes that the number of accident years is
equal to the number of observed development periods.}
\label{table:triangle}
\end{table}
}

\begin{landscape}
{\footnotesize
\begin{table}[ptbh]
\begin{center}
{\footnotesize {\scriptsize {\
\begin{tabular}{cc|cccccccccc|c}
\hline \multicolumn{1}{c|}{Parameters} & {Year} & {$0$} & {$1$} & {$2$} & {$3$} & {$4$} & {$5$} & {$6$} & {$7$} & {$8$} & {$9$} & {$\widehat{C}_{i,I}^{}-{C}_{i,I-i}$} \\
\hline \multicolumn{1}{c|}{$\left(\mathbf{f^{(CL)}}\right)$} & {$0$} & {$$} & {$$} & {$$} & {$$} & {$$} & {$$} & {$$} & {$$} & {$$} & \multicolumn{1}{c|}{$$} & {$0$} \\
\multicolumn{1}{c|}{$\left(\mathbf{f^{(MMSE)}}\right)$} & {$$} & {$$} & {$$} & {$$} & {$$} & {$$} & {$$} & {$$} & {$$} & {$$} & \multicolumn{1}{c|}{$$} & {$0$} \\
\hline \multicolumn{1}{c|}{$\left(\mathbf{f^{(CL)}}\right)$} & {$1$} & {$$} & {$$} & {$$} & {$$} & {$$} & {$$} & {$$} & {$$} & {$$} & \multicolumn{1}{c|}{$10,663,318$} & {$15,126$} \\
\multicolumn{1}{c|}{$\left(\mathbf{f^{(MMSE)}}\right)$} & {$$} & {$$} & {$$} & {$$} & {$$} & {$$} & {$$} & {$$} & {$$} & {$$} & \multicolumn{1}{c|}{$10,664,164$} & {$15,972$} \\

\hline \multicolumn{1}{c|}{$\left(\mathbf{f^{(CL)}}\right)$} & {$2$} & {$$} & {$$} & {$$} & {$$} & {$$} & {$$} & {$$} & {$$} & {$10,646,884$} & \multicolumn{1}{c|}{$10,662,008$} & {$26,257$} \\
\multicolumn{1}{c|}{$\left(\mathbf{f^{(MMSE)}}\right)$} & {$$} & {$$} & {$$} & {$$} & {$$} & {$$} & {$$} & {$$} & {$$} & {$10,645,322$} & \multicolumn{1}{c|}{$10,661,290$} & {$25,540$} \\

\hline \multicolumn{1}{c|}{$\left(\mathbf{f^{(CL)}}\right)$} & {$3$} & {$$} & {$$} & {$$} & {$$} & {$$} & {$$} & {$$} & {$9,734,574$} & {$9,744,764$} & \multicolumn{1}{c|}{$9,758,606$} & {$34,538$} \\
\multicolumn{1}{c|}{$\left(\mathbf{f^{(MMSE)}}\right)$} & {$$} & {$$} & {$$} & {$$} & {$$} & {$$} & {$$} & {$$} & {$9,736,710$} & {$9,745,473$} & \multicolumn{1}{c|}{$9,760,092$} & {$36,023$} \\

\hline \multicolumn{1}{c|}{$\left(\mathbf{f^{(CL)}}\right)$} & {$4$} & {$$} & {$$} & {$$} & {$$} & {$$} & {$$} & {$9,837,277$} & {$9,847,906$} & {$9,858,214$} & \multicolumn{1}{c|}{$9,872,218$} & {$85,302$} \\
\multicolumn{1}{c|}{$\left(\mathbf{f^{(MMSE)}}\right)$} & {$$} & {$$} & {$$} & {$$} & {$$} & {$$} & {$$} & {$9,840,743$} & {$9,853,536$} & {$9,862,404$} & \multicolumn{1}{c|}{$9,877,198$} & {$90,283$} \\

\hline \multicolumn{1}{c|}{$\left(\mathbf{f^{(CL)}}\right)$} & {$5$} & {$$} & {$$} & {$$} & {$$} & {$$} & {$10,005,044$} & {$10,056,528$} & {$10,067,393$} & {$10,077,931$} & \multicolumn{1}{c|}{$10,092,247$} & {$156,494$} \\
\multicolumn{1}{c|}{$\left(\mathbf{f^{(MMSE)}}\right)$} & {$$} & {$$} & {$$} & {$$} & {$$} & {$$} & {$10,019,212$} & {$10,074,318$} & {$10,087,415$} & {$10,096,493$} & \multicolumn{1}{c|}{$10,111,638$} & {$175,886$} \\

\hline \multicolumn{1}{c|}{$\left(\mathbf{f^{(CL)}}\right)$} & {$6$} & {$$} & {$$} & {$$} & {$$} & {$9,419,776$} & {$9,485,469$} & {$9,534,279$} & {$9,544,580$} & {$9,554,571$} & \multicolumn{1}{c|}{$9,568,143$} & {$286,121$} \\
\multicolumn{1}{c|}{$\left(\mathbf{f^{(MMSE)}}\right)$} & {$$} & {$$} & {$$} & {$$} & {$$} & {$9,422,181$} & {$9,501,327$} & {$9,553,584$} & {$9,566,004$} & {$9,574,613$} & \multicolumn{1}{c|}{$9,588,975$} & {$306,953$} \\

\hline \multicolumn{1}{c|}{$\left(\mathbf{f^{(CL)}}\right)$} & {$7$} & {$$} & {$$} & {$$} & {$8,445,057$} & {$8,570,389$} & {$8,630,159$} & {$8,674,568$} & {$8,683,940$} & {$8,693,030$} & \multicolumn{1}{c|}{$8,705,378$} & {$449,167$} \\
\multicolumn{1}{c|}{$\left(\mathbf{f^{(MMSE)}}\right)$} & {$$} & {$$} & {$$} & {$$} & {$8,448,582$} & {$8,576,155$} & {$8,648,195$} & {$8,695,760$} & {$8,707,065$} & {$8,714,901$} & \multicolumn{1}{c|}{$8,727,973$} & {$471,761$} \\

\hline \multicolumn{1}{c|}{$\left(\mathbf{f^{(CL)}}\right)$} & {$8$} & {$$} & {$$} & {$8,243,496$} & {$8,432,051$} & {$8,557,190$} & {$8,616,868$} & {$8,661,208$} & {$8,670,566$} & {$8,679,642$} & \multicolumn{1}{c|}{$8,691,971$} & {$1,043,242$} \\
\multicolumn{1}{c|}{$\left(\mathbf{f^{(MMSE)}}\right)$} & {$$} & {$$} & {$$} & {$8,229,268$} & {$8,421,009$} & {$8,548,167$} & {$8,619,971$} & {$8,667,381$} & {$8,678,649$} & {$8,686,460$} & \multicolumn{1}{c|}{$8,699,489$} & {$1,050,760$} \\

\hline \multicolumn{1}{c|}{$\left(\mathbf{f^{(CL)}}\right)$} & {$9$} & {$$} & {$8,470,989$} & {$9,129,696$} & {$9,338,521$} & {$9,477,113$} & {$9,543,206$} & {$9,592,313$} & {$9,602,676$} & {$9,612,728$} & \multicolumn{1}{c|}{$9,626,383$} & {$3,950,814$} \\
\multicolumn{1}{c|}{$\left(\mathbf{f^{(MMSE)}}\right)$} & {$$} & {$$} & {$8,477,596$} & {$9,121,045$} & {$9,333,566$} & {$9,474,503$} & {$9,554,088$} & {$9,606,636$} & {$9,619,125$} & {$9,627,782$} & \multicolumn{1}{c|}{$9,642,223$} & {$3,966,655$} \\

\hline{} & {$\widehat{f}_j^{(CL)}$} & {$1.4925$} & {$1.0778$} & {$1.0229$} & {$1.0148$} & {$1.0070$} & {$1.0051$} & {$1.0011$} & {$1.0010$} & {$1.0014$} & & {$6,047,061$}\\
{} & {$\sigma_j^{(CL)}$} & {$135.253$} & {$33.803$} & {$15.760$} & {$19.847$} & {$9.336$} & {$2.001$} & {$0.823$} & {$0.219$} & {$0.059$} & & \\
{} & {$\widehat{f}_j^{(MMSE)}$} & {$1.4937$} & {$1.0759$} & {$1.0233$} & {$1.0151$} & {$1.0084$} & {$1.0055$} & {$1.0013$} & {$1.0009$} & {$1.0015$} &  & {$6,139,834$}\\
{} & {$\sigma_j^{(MMSE)}$} & {$132.917$} & {$34.566$} & {$14.742$} & {$21.972$} & {$8.547$} & {$2.736$} & {$0.789$} & {$0.159$} & {$0.061$} & {$$}\\
\end{tabular}
} }  }
\end{center}
\caption{Predicted parameter estimates, $\widehat{\mathbf{f}},
\widehat{\bm{\sigma}}$, cumulative  chain ladder claims,
$\widehat{C}_{i,j}$, and estimated chain ladder reserves under the
classical ($CL$) and Bayesian ($MMSE$) models.}
\label{table:dfcl-predict}
\end{table}
}
\end{landscape}

\end{document}